\documentclass[twocolumn,showpacs,aps,prl]{revtex4}
\usepackage{graphics}
\usepackage{epsfig}
\begin{document}

\title{Ancilla-assisted quantum process tomography}
\author{J. B. Altepeter, D. Branning, E. Jeffrey, T. C. Wei, and P.
G. Kwiat}
\email{kwiat@uiuc.edu}
\affiliation{Dept. of Physics, University of Illinois at
Urbana-Champaign, Urbana IL 61801-3080}

\author{R. T. Thew}\thanks{Group of Applied Physics, University of Geneva, 1211 Geneva 4,
Switzerland}
\author{J. L. O'Brien}
\author{M. A. Nielsen}
\author{A. G. White} \email{andrew@physics.uq.edu.au}
\affiliation{Center for Quantum Computer Technology and School of
Physical Sciences, University of Queensland, QLD 4072, Brisbane,
Australia}

\date{March 7, 2003}

\begin{abstract} \noindent
Complete and precise characterization of a quantum dynamical
process can be achieved via the method of quantum process
tomography.  Using a source of correlated photons, we have
implemented several methods, each investigating a wide range of
processes, e.g., unitary, decohering, and polarizing.  One of
these methods, ancilla-assisted process tomography (AAPT), makes
use of an additional ``ancilla system,'' and we have theoretically
determined the conditions when AAPT is possible. 
Surprisingly, entanglement is \emph{not} required.
We present data obtained using 
both separable and entangled input states.  The use of 
entanglement yields superior results, however.
\end{abstract}

\maketitle

\vspace{-.2cm}
 Quantum information science~\cite{Nielsen00a}
exploits quantum mechanics to achieve information processing tasks
impossible in the classical world.  Recent
experiments~\cite{recexpt} have reported the implementation of a
wide variety of simple quantum information processing tasks. It is
important to benchmark the performance of experimental systems as
quantum information processing devices: one promising method,
proposed in 1997, is \emph{quantum process tomography}
(QPT)~\cite{Chuang97a}. Standard QPT (SQPT) involves preparing an
ensemble of a number of \emph{different} quantum states,
subjecting each of them to the (fixed) quantum process to be
characterized, and performing quantum state tomography on the
outputs. An alternative to SQPT, which we refer to as
\emph{ancilla-assisted process tomography} (AAPT), introduces an
extra \emph{ancilla} qubit, and involves preparation and
tomography of only a \emph{single} 2-qubit quantum state, rather
than four 1-qubit states~\cite{Leung}. As a special case,
entanglement-assisted process tomography (EAPT) describes the
situation when the ancilla is initially maximally entangled with
the system being characterized.

To date, SQPT has been realized in liquid nuclear magnetic
resonance systems~\cite{Nielsen98b} while SQPT and EAPT have been
demonstrated in optical systems, but only for unitary
transforms~\cite{d'Ariano02a}. Here we describe optical
implementations of SQPT, EAPT, and non-entangled AAPT for a
variety of processes, including unitary, decohering, and non-trace
preserving (e.g., partial polarizing) operations.  We also report
a theoretical result completely characterizing the class of states
usable for AAPT.  An equivalent result was independently developed,
and was reported just prior to our own~\cite{theoryResult}.

In SQPT, a quantum system, $A$, experiences an unknown quantum
process, ${\cal E}$.  To determine ${\cal E}$ we first choose some
fixed set of states $\{\rho_j\}$ which form a basis for the set of
operators acting on the state space of system $A$, e.g., $\{
\rho_j \}=\{\rho_H,\rho_V,\rho_D,\rho_R \}$ for a polarization
qubit (throughout this paper H, V, D, A, R, and L denote
Horizontal, Vertical, Diagonal, Anti-diagonal, Right-circular, and
Left-circular polarization, respectively). Each state $\rho_j$ is
then subject to the process ${\cal E}$, and quantum state
tomography \cite{tomoref1,tomoref2,tomoref3} is used to
experimentally determine the output ${\cal E}(\rho_j)$.  ${\cal
E}$ is fully characterized if we determine matrices $E_j$, known
as \emph{operation elements}, such that ${\cal E}(\rho) = \sum_j
E_j \rho E_j^{\dagger}, \ \forall \rho$. This representation is
known as an operator-sum decomposition~\cite{Nielsen00a}.

In AAPT the process ${\cal E}$ is characterized by preparing a
\emph{single state}, $\sigma$, and then measuring $({\cal E}
\otimes {\cal I})(\sigma)$. This requires an \emph{ancilla
system}, $B$, with Hilbert space dimension at least as great as
that of $A$.  For an appropriate initial state, it is possible to
characterize ${\cal E}$ by preparing the state $\sigma$,
performing the process ${\cal E}$ on system $A$
--- leaving system $B$ completely isolated --- and taking a
tomography of the output $({\cal E} \otimes {\cal I})(\sigma)$.
The total number of measurements is the same in AAPT (16
measurements on a single 2-qubit state) as in SQPT (four
measurements on each of four input states).

AAPT has advantages over SQPT, most notably being that preparation
of only a single quantum state is necessary for its operation.
Consider the possibility of using it as a diagnostic tool in a
quantum computer.  When an unknown effect acts on less than half
of a system of qubits, knowledge of the larger state before and
after the change is sufficient to exactly predict the effect this
change will have on every other state. (Assuming that the larger
state is usable for AAPT --- see below). Alternatively, SQPT has
the advantage that it is generally easier to produce and measure
states with fewer qubits.

We have investigated a variety of dynamical processes, using the
three methods of SQPT, EAPT, and non-entangled AAPT. Our processes
operate on the polarization state of a single photon. We used
spontaneous parametric downconversion (of a 351-nm pump beam) in a
nonlinear crystal (BBO) to create pairs of time-correlated photons
at 702 nm.  For SQPT, by triggering on one photon, the other was
prepared into a single-photon state~\cite{mandel} with H
polarization (Fig. 1).  Half and quarter waveplates converted the
horizontal polarization into an arbitrary state, thus allowing us
to prepare the necessary input states $\rho_H,\rho_V,\rho_D,$ and
$\rho_R$. The tomography of the post-process states was performed
by measuring (in coincidence with the trigger detector) the Stokes
parameters $S_{1} = P_H - P_V$, $S_{2} = P_D - P_A$, and $S_{3} =
P_R - P_L$, and performing a maximum-likelihood estimation of the
density matrix~\cite{tomoref2}. (Here $P_i$ denotes a
\emph{probability}: calculated as the intensity of a state
measured in the $i^{th}$ basis divided by the total intensity.)
Typical measurements yielded a maximum of 13,000 photon counts
over 30 seconds.

For our EAPT results, two adjacent BBO crystals were used to
prepare the maximally entangled state $|\phi^{-} \rangle =
(|HH\rangle - |VV\rangle)/\sqrt{2})$ \cite{footnote4}.  One of the
resulting qubits was subjected to the given process, and two-qubit
tomography of the pair was then performed by measuring the
polarization correlations of the photons with 16 measurements,
e.g., in the following bases: HH, HV, HD, HR, VH, VV,
etc.~\cite{tomoref1}. Note from Fig.~1 that the elements used in
SQPT to prepare the single photon state are now placed (in reverse
order) in the other detection arm, highlighting the symmetry of
the two techniques.

\begin{figure}[t]
\begin{center}
\epsfig{file=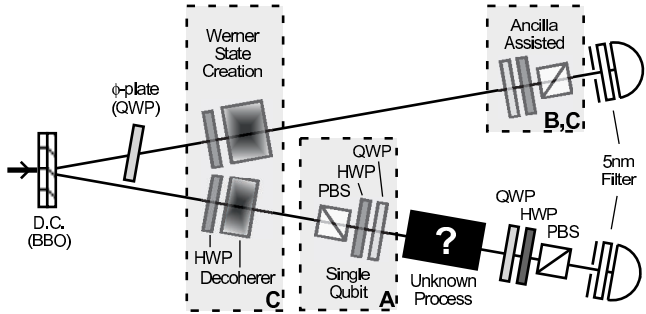, width =8.5cm}
\end{center}
\vspace{-.3cm} \footnotesize \caption{Experimental arrangements to
perform quantum process tomography. A 351-nm pump is directed
through two 0.6mm-thick BBO crystals, giving rise to pairs of
correlated photons at 702 nm, which are detected using Si
avalanche photodiodes and fast coincidence electronics. A, B, and
C above denote which elements are present for SQPT, EAPT, and
non-entangled AAPT, respectively. a.) Single-qubit process
tomography: Polarizer (P), half waveplate (HWP) and quarter
waveplate (QWP) allow preparation of required pure single-photon
(conditioned on ``trigger'' detection) states; identical elements
allow tomography of the post-process states. b.)
Entanglement-assisted tomography: The source produces the
maximally entangled state $(|HH \rangle - |VV \rangle)/\sqrt{2}$.
A two-photon tomography of the output allows reconstruction of the
process. c.) Ancilla-assisted tomography: The source produces the
Werner state $\rho_{W} \sim \frac{1}{6}I + \frac{1}{3}| \gamma
\rangle \langle \gamma|)$, where $| \gamma \rangle$ is a maximally
entangled state. Although there is no entanglement, the
correlations in $\rho_{W}$ allow AAPT. } \vspace{-.2 cm}
\end{figure}

We also performed AAPT using the non-entangled Werner state
$\rho_{W} = {\frac{1}{6}}I + {\frac{1}{3}}| \gamma \rangle \langle
\gamma|$, where $| \gamma \rangle$ is a maximally entangled state.
To prepare this state we adjust the polarization of the pump beam
until the down conversion crystals produce the pure, partially
entangled state $\frac{1}{\sqrt{6}(\sqrt{2}-1)}|HH\rangle +
\frac{\sqrt{2}-1}{\sqrt{6}}|VV\rangle$~\cite{tomoref1}. A half
waveplate at $22.5^\circ$ in each arm then transforms this state
into $| \phi \rangle = \sqrt{\frac{1}{3}}|HH \rangle +
\sqrt{\frac{1}{6}}|HV \rangle + \sqrt{\frac{1}{6}}|VH \rangle +
\sqrt{\frac{1}{3}}|VV \rangle$. Next we pass each photon through a
decoherer, an 11-cm piece of quartz which separates the horizontal
and vertical components of the polarization by $\sim$ 100~$\mu$m,
which is the coherence length of the individual photons
(determined by the 3-mm diameter collection irises and the 5-nm
bandwidth (FWHM) interference filters.)  This destroys all
coherence terms in $| \phi \rangle \langle \phi|$ except for $| HH
\rangle \langle VV|$ and $| VV \rangle \langle HH|$. An
additional, shorter decoherer in the idler arm lowers these terms
to achieve a state which has $ 99.2 \pm .8\% $
fidelity~\cite{fidelity} with the above Werner state.

For single-qubit processes, a convenient graphical representation
plots the transformation of the sphere of all possible states
(e.g., the Poincar\'{e} sphere for
polarization)~\cite{Nielsen00a}, as determined by the action of
the process on the set of basis states, $\rho_{j}$. For example,
all unitary transformations are equivalent to a rotation about
some axis (Fig.~2b). Decoherence is represented by a collapsing of
the sphere toward a ``spindle'' (Fig.~2c); for instance, complete
decoherence in the HV basis leaves the states $|H\rangle$ and
$|V\rangle$ unmodified, but transforms the states $|D\rangle$ and
$|R\rangle$ into the completely mixed state at the center of the
sphere. This graphical approach can even be applied to lossy
processes, e.g., partial polarizers, though it is important to
note that it does not indicate the amount of loss, only the
quantum state of the surviving qubits.

\begin{figure}[t]
\begin{center}
\epsfig{file=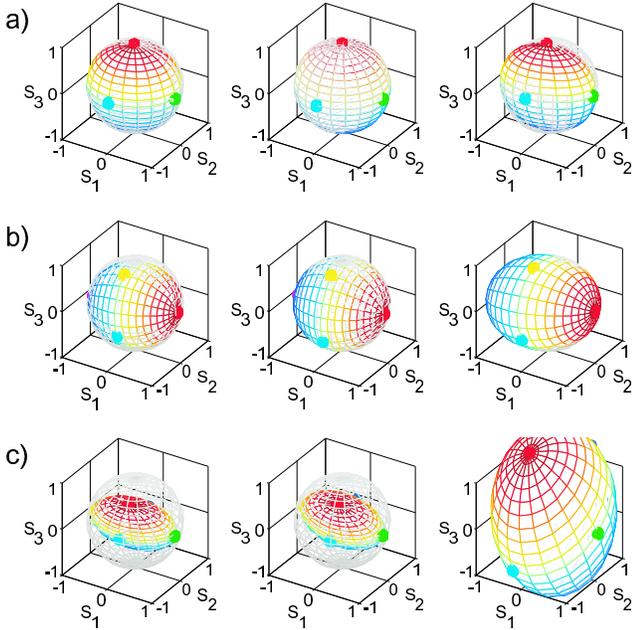, width =8.5cm}
\end{center}
    \vspace{-.6cm} \footnotesize \caption{Geometric mappings for three
    quantum processes -- (a) identity , (b) unitary transformation,
    and (c) decoherence  -- measured using SQPT (left),
    EAPT (center), and AAPT (right).  The axes are the Stokes parameters ($S_{1}$,
    $S_{2}$, $S_{3}$).  The colored mesh surfaces show how all pure
    states are transformed by the process.  The initial
    states H, R, V, and A are shown by the green, red,
    yellow, and blue dots, respectively.  The transformation of initial {\it mixed}
    states (inside the surface) may be interpolated from the pure state
    results using the linearity of quantum mechanics.  The mesh
    coloring denotes the orientation of the transformed sphere.
    }
\vspace{-.5 cm}
\end{figure}

We now outline the general procedure for SQPT, as described
in~\cite{Chuang97a}. Rather than directly determining the
operation elements ${ E_{j}}$, SQPT relates these to a fixed set
of operators, $\{ \widetilde{E}_{m} \}$, where $E_{j} = \sum_{m}
e_{jm} \widetilde{E}_{m}$ and $e_{jm}$ can be
complex.  This allows us to define a \emph{single} 
matrix, $\chi$, that fully characterizes the process: if we
rewrite the process as ${\cal E}(\rho) = \sum_{mn}
\widetilde{E}_{m} \rho \widetilde{E}_{n}^{\dagger} \chi_{mn}$ then
$\chi$ is a positive Hermitian matrix, $\chi_{mn} = \sum_k
e_{km}e_{kn}^*$. See Fig.~3 for examples of experimentally
determined $\chi$ matrices.  To determine $\chi$ from a set of
measurements, we choose a set of basis states $\{ \rho_j \}$, such
that for each input state ${\rho}_{j}$, state tomography returns
an output, ${\cal E} (\rho_{j}) = \sum_{k}c_{jk}\rho_{k}$. If we
define $\widetilde{E}_{m} \rho_{j} \widetilde{E}_{n}^{\dagger} =
\sum_{k} \beta_{jk}^{mn} \rho_{k}$ (where $\beta_{jk}^{mn}$ is
another complex number matrix which we determine from our choice
of input basis states \{$\rho_j$\}, output basis states
\{$\rho_k$\},
 and operators \{$\widetilde{E}_j$\}), we can
see that $\sum_{k} \sum_{mn} \chi_{mn} \beta_{jk}^{mn} \rho_{k} =
\sum_{k} c_{jk} \rho_{k}$, independent of $\rho_{k}$; $\beta$ is
invertible; and $\chi_{mn} = \sum_{jk} \left( \beta^{-1}
\right)_{jk}^{mn} c_{jk}$.

\begin{figure}[b]
\begin{center}
\includegraphics[width=8.5cm]{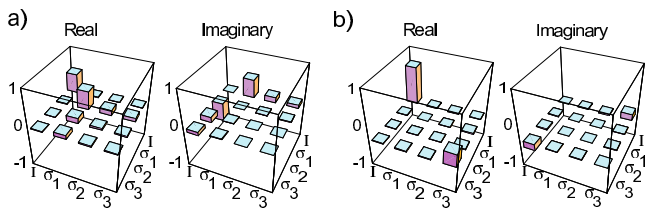}
\end{center}
\vspace{-.5cm} \footnotesize \caption{$\chi$-matrices determined
from EAPT for (a) unitary and (b) decohering processes, as shown
in Fig.  2. }
\end{figure}

In our experiment we use $\{ \widetilde{E}_{m} \}=\{ I, \sigma_x,
\sigma_y, \sigma_z \}$, respectively equivalent to the following
optical elements: nothing; a HWP plate at 45$^{\circ}$; an
optically active element; a HWP plate at 0$^{\circ}$. The diagonal
elements of the $\chi$-matrix correspond, respectively, to the
probability of carrying out the $I, \sigma_x, \sigma_y,$ and
$\sigma_z$ processes, while the off-diagonal elements correspond
to \emph{coherence} processes of the form $ \sigma_x \rho
\sigma_y$ and $ \sigma_y \rho \sigma_x$, etc.

We investigated several processes, including the identity, a
unitary rotation, a decoherer, and both a coherent and an
incoherent partial polarizer (see below).  The results for the
identity process measure how well the input state(s) are
preserved.  We used SQPT, EAPT, and AAPT to measure the same
unitary rotation process (a birefringent waveplate). The results
were in close agreement (Fig.~2b); the resulting $\chi$ matrices
had an average process fidelity \cite{processFidelity} between the
three methods, of ${\cal F} = 100.4\pm .8\% $. Likewise, the SQPT
and EAPT measurements of a decohering process (implemented with a
6.3-mm piece of quartz) yielded $ {\cal F} =99.9\pm .3\% $
(Fig.~2c). The same process, when measured using our Werner State,
appears to be a {\it recoherer}
--- a process which is not a positive map.  Recall
that this Werner state was prepared using a thick piece of quartz
to temporally separate the H and V components of the light,
introducing decoherence.  Consider adding another piece of quartz,
with optic axis perpendicular to the first, after the original.
This also temporally shifts the H and V components of the light,
but in the opposite direction, undoing the original decoherence.
Our decohering process does exactly this, effectively {\it
recohering} the Werner State --- impossible for a 1-qubit process.
The resolution to this paradox lies in the implicit assumption
that the measured process does not act on any degrees of freedom
used to prepare the input state other than the tested qubit.  For
example, if frequency is traced over to prepare a mixed input
state, a process that couples to frequency cannot be measured.

Coherent and incoherent partial polarizers were measured in order
to highlight the role coherence plays in lossy processes. A glass
plate at Brewster's angle to an incident beam is a coherent
partial polarizer, as the operation of the plate maintains the
pre-existing phase relationship between the horizontal component
of the light (completely transmitted) and the vertical component
of the light (partially reflected).  For the incoherent case,
consider inserting a horizontal polarizer into the beam $50\%$ of
the time. Half the time only the horizontal component of the light
will be transmitted, but more importantly, the transmitted light
will have no coherence relationship with the light that does not
pass through the polarizer.  For the coherent partial polarizer,
pure states remain pure but slide toward H along the surface of
the sphere. In the incoherent case pure states travel linearly
through the sphere to H, becoming mixed (Fig.~4).

\begin{figure}[t]
\begin{center}
\includegraphics[width=7cm, bb=0 0 193 200]{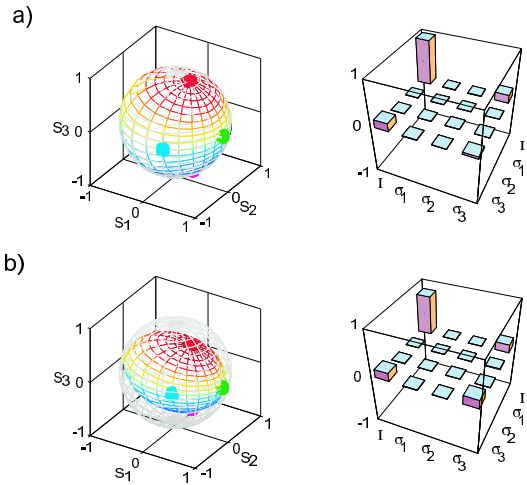}
\end{center}
\vspace{-.6cm} \footnotesize \caption{Geometric mappings and
$\chi$ matrices for (a) coherent and (b) incoherent partially
polarizing processes.  The former was implemented using two glass
plates (microscope slides) near Brewster's angle [$T_{H} \sim
88\%, T_{V} \sim 45\%$]. The latter was simulated by inserting a
horizontal polarizer 50\% of the time.  (Real components shown;
imaginary contributions $< 1\%$.) } \vspace{-.3cm}
\end{figure}

What class of initial states $\sigma$ of the $AB$ system may be used
for AAPT? This question can be answered using an operator
generalization of the Schmidt decomposition for entangled
states~\cite{Nielsen00a}. To explain this decomposition, we introduce
an inner product on operators, $(M,N) \equiv \mbox{tr} (M^ \dagger
N)$, and define an orthonormal operator basis to be a set of operators
$\{ M_j \}$ such that $(M_j,M_k) = \mbox{tr} (M_j^\dagger M_k) =
\delta_{jk}$.  (For example, an orthonormal basis for single-qubit
operators is the set $\{ I/\sqrt 2, \sigma_x/\sqrt 2, \sigma_y/\sqrt
2, \sigma_z/\sqrt 2 \}$). The operator-Schmidt
decomposition~\cite{Nielsen98d} states that an operator $M$ acting on
$AB$ can be decomposed as $M = \sum_l s_l A_l \otimes B_l$, where the
$s_l$ are non-negative real numbers, and the sets $\{ A_l \}$ and $\{
B_l \}$ form orthonormal operator bases for systems $A$ and $B$,
respectively \cite{foot3}. The \emph{Schmidt number} $\mbox{Sch}(M)$
of an operator $M$ is defined \cite{Nielsen98d} as the number of
non-zero terms in the Schmidt decomposition.

A state $\sigma$ of $AB$ may be used to perform AAPT \emph{if and
only if} the Schmidt number of $\sigma$ is $d_A^2$, where $d_A$ is
the dimension of the state space of system $A$.  Consider that in
order to measure the mapping of the entire space, the input state
must possess correlations - represented by the Schmidt number -
between enough states to form a basis for the mapping.  To prove
this, expand $\sigma$ in its Schmidt decomposition as $\sigma =
\sum_l s_l A_l \otimes B_l$. Assume $\sigma$ has Schmidt number
$d_A^2$, so that the $A_l$ form an orthonormal operator basis, and
$s_l > 0$ for all $l$. Let $\sigma'$ be the output obtained after
letting ${\cal E}$ act on system $A$, that is, $\sigma' = ({\cal
E} \otimes {\cal I})(\sigma) = \sum_l s_l {\cal E}(A_l) \otimes
B_l$. By the orthonormality of the $B_l$ and the previous equation
it follows that $\mbox{tr}_B( (I \otimes B_m^\dagger) \sigma') =
\sum_l s_l {\cal E}(A_l) \mbox{tr} (B_m^\dagger B_l) = s_m {\cal
E} (A_m)$, and so ${\cal E}(A_m) = \mbox{tr}_B((I\otimes
B_m^\dagger)\sigma')/s_m$.  The fact that the Schmidt number of
$\sigma$ is $d_A^2$ ensures that $s_m > 0$, so there is no problem
with division by zero.  It follows that it is possible to
determine the action of ${\cal E}$ on an operator basis by doing
state tomography on $\sigma'$, and applying the above equation.
The techniques described earlier can then be used to generate a
$\chi$ matrix or transformed sphere.

Conversely, let $E_A$ be the space of trace-preserving quantum
operations on system $A$, and let $S_{AB}$ be the space of quantum
states on system $AB$. Define a map $f:E_A \rightarrow S_{AB}$ by
$f({\cal E}) \equiv ({\cal E} \otimes {\cal I})(\sigma)$.  For
AAPT, we require that $f$ be a one-to-one map, i.e., there are
never two distinct operations such that $f({\cal E}_1) =
f({\cal E}_2)$.  A parameter counting argument shows that $f$
cannot be one-to-one when $\sigma$ has Schmidt number less than
$d_A^2$. The dimensionality of the manifold $E_A$ is
$d_A^4-d_A^2$. Since $f({\cal E}) = \sum_l s_l {\cal E}(A_l)
\otimes B_l$, the dimension of the image manifold $f(E_A)$ is at
most $\mbox{Sch}(M) \times (d_A^2-1)$, because the map ${\cal E}
\rightarrow {\cal E}(A_l)$ has image of dimension at most
$d_A^2-1$. Thus, for AAPT we require that $\mbox{Sch}(M) \times
(d_A^2-1) \geq d_A^4-d_A^2$, which is only possible when
$\mbox{Sch}(M) = d_A^2$.

Note that AAPT is possible only when the dimension of system $B$
is at least as great as the dimension of system $A$.  When this is
true, almost all states of system $AB$ may be used for AAPT,
because the set of states with Schmidt number less than $d_A^2$
has measure zero. That is, a maximally entangled input is not
required for AAPT --- indeed many of the viable input states are
not entangled at all, as demonstrated by our Werner state AAPT.
However, while almost any state \emph{can} be used for AAPT,
maximally entangled states appear to be experimentally optimal in
that they have perfect non-local correlations.  Fig.~2 highlights
this difference, as the AAPT results have significantly greater
statistical errors than the EAPT (both were from identical
measurement runs).  This comparative usefulness of entangled versus 
separable states was first introduced and is discussed further in
\cite{theoryResult}.

\noindent We would like to acknowledge partial support from the
NSF (\#EIA-0121568) and the DCI Postdoctoral Research Fellowship
Program.  We would like to thank Debbie Leung for helpful
conversations.


\begin{thebibliography}{99}
\bibitem{fidelity} $F(\rho_1, \rho_2) =
(Tr\sqrt{\sqrt{\rho_1}\rho_2\sqrt{\rho_1}})^2$, which - for
$\rho_1$ or $\rho_2$ pure - simplifies to $Tr(\rho_1\rho_2)$ [2].
\end{thebibliography}
\end{document}